\documentclass[runningheads]{llncs}
\usepackage[english]{babel}
\usepackage{graphicx}
\usepackage{amsmath}
\usepackage{amsfonts}
\usepackage{amssymb}

\def\B{\mathcal{B}}

\def\nontrans{\texttt{Non-transitive}}
\def\polish{\texttt{Transitive-predefined}}
\def\russianT{\texttt{Transitive-ab-initio}}
\def\russianNT{\texttt{Non-tree-transitive}}

\def\q#1{\mathtt{#1}}

\def\1{\q{1}}
\def\0{\q{0}}

\newcommand{\al}{\alpha}

\newcommand{\G}{\mathcal{B}}

\title{Efficient seeding techniques for \newline protein similarity search}
\titlerunning{subset seeds for protein similarity search}
\author{
Mikhail Roytberg\inst{1}
\and
Anna Gambin\inst{2}
\and
Laurent No{\'e}\inst{3}
\and
S\l{}awomir Lasota\inst{2}
\and
\newline
Eugenia Furletova\inst{1} 
\and
Ewa Szczurek\inst{4}
\and
Gregory Kucherov\inst{3}
}
\institute{
Institute of Mathematical Problems in Biology, Pushchino, Moscow
Region, 142290, Russia, \email{mroytberg@mail.ru,furletova@impb.psn.ru}
\and
Institute of Informatics, Warsaw University, Banacha 2, 02-097, Poland, \email{\{aniag|S.Lasota\}@mimuw.edu.pl}
\and
LIFL/CNRS/INRIA, B\^at. M3, Campus Scientifique,
59655 Villeneuve d'Ascq C\'edex, France, \email{\{Gregory.Kucherov|Laurent.Noe\}@lifl.fr}
\and
Max Planck Institute for Molecular Genetics,
Computational Molecular Biology,
Ihnestr. 73,
14195 Berlin, Germany, \email{ewa.szczurek@molgen.mpg.de}
}
\authorrunning{M.~Roytberg, A.~Gambin, L.~No{\'e}, S.~Lasota,  E.~Szczurek, G.~Kucherov}

\date{{\normalsize{\today}}}

\begin{document}
\maketitle
\begin{abstract}
We apply the concept of {\em subset seeds} proposed in
\cite{KucherovNoeRoytberg06} to similarity search in protein
sequences. The main question studied is the design of efficient 
{\em seed alphabets} to construct seeds with optimal
sensitivity/selectivity trade-offs. We propose several different
design methods and use them to construct several alphabets. We then
perform an analysis of seeds built over those alphabet and
compare them with the standard {\sc Blastp} seeding method
\cite{BLAST90,GBLAST97}, as well as 
with the family of vector seeds proposed in \cite{BrownTCBB05}. While
the formalism of subset seed is less expressive (but less costly to
implement) than the accumulative
principle used in {\sc Blastp} and vector seeds, our seeds show a
similar or even better performance than {\sc Blastp} on Bernoulli
models of proteins compatible with the common BLOSUM62
matrix.
\end{abstract}
\section{Introduction}

Similarity search in protein sequences is probably the most classical
bioinformatics problem, and a commonly used algorithmic solution is
implemented in the ubiquitous {\sc Blast} software
\cite{BLAST90,GBLAST97}. On the other hand, similarity search
algorithms for nucleotide sequences (DNA, RNA) underwent
several years ago a significant improvement due to the idea of 
{\em spaced seeds} and its various generalizations
\cite{PatternHunter02,PatternHunter04,BrejovaBrownVinarJCSS05,NoeKucherovNAR05,MakGelfandBensonBioinformatics06,CsurosMaAlgorithmica07,ZhouStantonFloreaBMCBioinformatics08}.
This development, however, has little affected protein sequence
comparison, although improving the speed/precision trade-off for protein
search would be of great value for numerous bioinformatics projects. 
Due to a bigger alphabet
size, protein seeds are much shorter (typically 2-5 letters instead of
10-20 letters in the DNA case) and also letter identity is much less
relevant in defining hits than in the DNA case. For these reasons, the
spaced seeds technique might seem not to apply directly to protein
sequence comparison. 

Recall that {\sc Blast} applies quite different approaches to protein
and DNA sequences to define a hit. In the DNA case, a hit is defined as a
short pattern of identically matching nucleotides whereas in the
protein case, a hit is defined through an {\em accumulative}
contribution of a few amino acid matches (not necessarily identities) 
using a given {\em scoring matrix}. 
Defining a hit through an additive contribution of several positions
is captured by a general formalism of {\em vector seeds} proposed in
\cite{BrejovaBrownVinarJCSS05}. On the other hand, it has been
understood
\cite{PatternHunter04,BuhlerRECOMB04,KucherovNoeRoytberg04,YangWangChenEtAlBIBE04,XuBrownLiMaCPM04}
that using simultaneously a {\em family} of seeds instead of a single
seed can further improve the sensitivity/selectivity ratio. Papers
\cite{BrownTCBB05,tPatternHunter05} both propose solutions using
a family of vector seeds to surpass the performance of {\sc Blast}. 

However, using the principle of accumulative score over several
adjacent positions has an algorithmic cost. 
Defining a hit through a pattern of exact letter matches
allows for a {\em direct hashing} scheme, where each key of the query
sequence is associated with a {\em unique} hash table entry storing
positions of the subject sequence (database) where the key can hit. On
the other hand, defining a hit through an accumulative contribution of
several positions leads to an additional pre-computed table storing,
for each key, its {\em neighborhood} i.e., the list of subject keys
(or corresponding hash table entries) with which it can form a
hit. For example, in a standard {\sc Blastp} setting (Blosum62 scoring
matrix with threshold 11 for accumulative score of 3 contiguous
positions) a 3-letter key hits on average 19.34 distinct keys,
i.e. requires that many accesses to the hash table. For the family of
vector seeds from \cite{BrownTCBB05} with an equivalent selectivity
level (score 18), a (here 4-letter) key hits on average 15.99 keys. 
For some applications, for example in setting large-scale protein
comparisons on a specialized computer architecture (see
e.g. \cite{PeterlongoEtAlPBC07}) one might need to minimize the number
of hash table accesses, and therefore to use another seeding
formalism. 

In \cite{KucherovNoeRoytberg06}, we proposed a new concept of 
{\em subset seeds} that can be viewed as an intermediate between
ordinary spaced seeds and vector seeds: subset seeds allow one to
distinguish between different types of mismatches (or matches) but
still treat seed positions independently rather than
cumulatively. Distinguishing different mismatches is not done by
scoring them, but by extending the seed alphabet such that each seed
letter specifies different sets of mismatches. For example, in the DNA
case it is beneficial to distinguish between transition mutations
({\tt A} $\leftrightarrow$ {\tt G}, {\tt C} $\leftrightarrow$ {\tt T})
and others (transversions) \cite{NoeKucherovBMC04}. 

Since the protein alphabet is much larger than the one of DNA, subset
seeds provide a very attractive seeding option for protein
alignment. The present study is then motivated by following
general questions: {\em how far can we go with subset seeds applied to
protein sequences? Can we reach the performance of {\sc Blast} seeds?
the one of vector seeds? or maybe even outperform them? \ldots}

In the paradigm of subset seeds, each seed letter specifies a set of
amino acid pairs matched by this letter. Therefore, a crucial question
is the design of an appropriate {\em seed alphabet}, which is one of
the main problems we study in this paper. In
Section~\ref{section:preliminaries}, we introduce some probabilistic
notions we need to reason about seed
efficiency. Section~\ref{section:non-transitive-seed-alphabet}
introduces the first simple approach to design a seed alphabet, which,
however, does not lead to so-called {\em transitive} seeds, useful in
practice. Section~\ref{section:transitive-seed-alphabet} presents
three 
different approaches to designing transitive seed alphabets, based on
a pre-defined (Section~\ref{subsection:prebuilded-tree}) or newly
designed (Section~\ref{subsection:ab-initio-clustering}) hierarchical
clustering of amino acids, as well as on a non-hierarchical clustering
(Section~\ref{subsection:non-hierarchical}).
Section~\ref{section:experiments} describes comparative experiments
made with the designed seeds on probabilistic models.

\section{Preliminaries}
\label{section:preliminaries}
Throughout the paper, $\Sigma=\{\mathtt{A, C, D, E, F, G, H, I,
  K, L, M, N, P, Q, R, S, T, V, W, Y}\}$ denotes the alphabet of amino acids.

In most general terms, a {\em (subset) seed letter} $\alpha$ is
defined as any symmetric and reflexive binary relation on 
$\Sigma$. Let $\B$ be a {\em seed alphabet}, i.e. a collection of 
subset seed letters. Then a {\em subset seed} $\pi=\alpha_1 \ldots
\alpha_k$ is a word over $\B$. $\pi$ defines a 
symmetric and reflexive binary relation on words of $\Sigma^k$ (called
{\em keys}): for 
$s_1,s_2\in \Sigma^k$, $s_1 \sim_\pi s_2$ iff $\forall i\in [1..k]$, we have
$\langle s_1[i],s_2[i]\rangle\in \alpha_i$. 

For practical reasons, we would like seed letters to define a 
{\em transitive} relation, in addition. This induces an equivalence
relation on keys, which is very convenient and allows for an efficient
indexing scheme (see Introduction). In this paper, we will be mainly
interested in transitive seed letters, but we will also study
the non-transitive case in order to see how restrictive the
transitivity condition is. 

The quality of a seed letter or of a seed is characterized by two main
parameters:  
{\em sensitivity} and {\em selectivity}. They are defined through
background and foreground probabilistic models of protein alignments.
Foreground probabilities are assumed to represent the distribution of
amino acids matches in proteins of interest, when two homologous
proteins are aligned together. Background probabilities, on the other
hand, represent the distribution of amino acid matches in 
{\em random alignments}, when two proteins are randomly aligned
together. 

In this paper, we restrict ourselves to Bernoulli models of proteins
and protein alignments, although some of
the results we will present can be extended to Markov models. 

Assume that we are given background probabilities
$\{b_1,\ldots,b_{20}\}$ of amino acids in protein sequences under
interest. The {\em background probability} of a seed letter
$\alpha$ is defined by $b(\alpha)=\sum_{(a_i, a_j)\in \alpha} b_i
b_j$. The {\em selectivity} of $\alpha$ is $1-b(\alpha)$ and the 
{\em weight} of $\alpha$ is defined by 
\begin{equation}
w(\alpha)=\frac{\log b(\alpha)}{\log b(\#)},
\end{equation}
where $\#=\{\langle a,a\rangle|a\in\Sigma\}$ is the
``identity'' seed letter.  
For a seed $\pi=\alpha_1 \ldots \alpha_k$, the background probability
of $\pi$ is $b(\pi)=\prod_{i=1}^k b(\alpha_i)$, the selectivity of
$\pi$ is $1-b(\pi)$ and the weight of $\pi$ is $w(\pi)=\log_{b(\#)}
  b(\pi)=\sum_{i=1}^k w(\alpha_i)$. 
Note that the weight here generalizes the weight of of classical
spaced seeds \cite{KeichLiMaTrompDAM04} defined as the number of
``identity'' letters it contains. 

Let $f_{ij}$ be the probability to see a pair $\langle a_i,a_j\rangle$ 
aligned in a target alignment. The {\em foreground probability} of a
seed letter $\alpha$ is defined by $f(\alpha)=\sum_{(a_i, a_j)\in \alpha} f_{ij}$. 
The {\em sensitivity} of a seed $\pi$ is defined as
the probability to hit a random target alignment\footnote{Note that our definitions of sensitivity and selectivity are not symmetric: sensitivity is
  defined with respect to the entire alignment and selectivity with respect to a
single alignment position. These definitions capture better the
intended parameters we want to measure. However, selectivity could
also be defined with respect to the entire alignment. We could suggest
the term {\em specificity} for this latter definition.}. 
Assume that target alignments are specified
by a length $N$. Then the sensitivity of a seed
$\pi=\alpha_1 \ldots\alpha_k$ is
the probability that a randomly drawn gapless alignment (i.e. string of
pairs $\langle a_i,a_j\rangle$) of length $N$ contains a fragment of length $k$
which is matched by $\pi$. In \cite{KucherovNoeRoytberg06} we proposed
a general algorithm to efficiently compute the seed sensitivity for a 
broad class of target alignment models. This algorithm will be used
in the experimental part of this work. 

The general problem of seed design is to obtain seeds with good
sensitivity/selectivity trade-off. Even within a fixed seed formalism, 
the quality of a seed is dependent on the chosen selectivity value. 
This is why we will always be interested in
computing efficient seeds for a large range of selectivity levels. 

\section{Dominating seed letters}
\label{section:non-transitive-seed-alphabet}
Our main question is how to choose seed letters that form good seeds?
Intuitively, ``good letters'' are those that best distinguish
foreground and background letter alignments. 

For each letter $\alpha$, consider its foreground and background
probabilities $f(\alpha)$ and $b(\alpha)$ respectively. 
Intuitively, we would like to have
letters $\alpha$ with large $f(\alpha)$ and small $b(\alpha)$. A
letter $\alpha$ is said to {\em dominate} a letter $\beta$ if $f(\alpha)\geq
f(\beta)$ and $b(\alpha)\leq b(\beta)$. Observe that in this case,
$\beta$ can be removed from consideration, as it can be always
advantageously replaced by $\alpha$. 

Consider all amino acid pairs $(a_i,a_j)$ ordered by descending 
{\em likelihood ratio} $f_{ij}/b_i b_j$. Consider the set of pairs
$(a_i,a_j)$ such that $f_{ij}/b_i b_j>s$ for some threshold value
$s$. Then one can show that this set forms a letter that cannot be
dominated by any other letter\footnote{It is interesting to
point out the relationship to the well-known Neyman-Pearson lemma which is a more
general formulation of this statement.} (proof omitted).
This observation leads to defining seed letters that consist of those
pairs $(a_i,a_j)$ for which the ratio $f_{ij}/b_i b_j$ is above a
given threshold.
\subsubsection{Resulting alphabet} 
We computed the likelihood ratio for all amino acid pairs, based on
practical values of 
background and foreground probabilities computed in
accordance with the BLOSUM62 matrix (see Section~\ref{subsection:methods}). 
Not surprisingly, amino acid identities (pairs $\langle a,a\rangle$)
have highest likelihood scores varying from $38.11$ for tryptophan
down to $3.69$ for valine. 
\begin{figure}[htb]\center
\vspace{-0.0cm}
\includegraphics[width=0.52\textwidth]{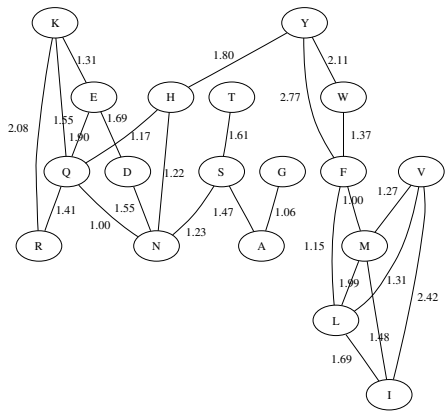}
\vspace{-0.1cm}
\caption{Alphabet {\nontrans}: amino acid pairs with likelihood ratio $> 1$}
\label{figure:likelihood_threshold}
\vspace{-0.0cm}
\end{figure}
Among distinct pairs, only $25$ have a score greater than $1$
(Figure~\ref{figure:likelihood_threshold}). A quick
analysis shows that those do
not form transitive relations, and therefore do not verify the
transitivity requirement. The alphabet containing
those 25 pairs is denoted {\nontrans}. It will be used in the
experimental part of the paper 
(Section~\ref{section:experiments}) in order to study how restrictive is the requirement of
transitive letters, i.e. how much better are general seeds than those
obtained with the restriction of transitivity. 

\section{Transitive seed alphabets}
\label{section:transitive-seed-alphabet}
In the case of transitive seed alphabets, every letter
$\al\in\B$ is a partition of the amino acid alphabet $\Sigma$. 
In other words, the binary relation associated with each letter 
(cf Section~\ref{section:preliminaries}) is an equivalence relation. 
Transitive alphabets represent the practical case when each amino acid 
is uniquely mapped to its equivalence class. This, in turn, allows
for an efficient hashing scheme during the stage of seed search, when 
different entries of the hash table index non-intersecting subsets of
keys. 

In Sections~\ref{subsection:prebuilded-tree},\ref{subsection:ab-initio-clustering},
we explore transitive seed alphabets verifying an additional  
condition: for any two seed letters $\al_1,\al_2\in\B$ corresponding to 
partitions $P_{\alpha_1},P_{\alpha_2}$ respectively, one of
$P_{\alpha_1},P_{\alpha_2}$ is a refinement of the other. Formally,
for any $\al_1,\al_2\in\B$, 
\begin{equation}
\mbox{ either every set } \sigma\in P_{\alpha_1}\mbox{ is a subset of
  some }\delta\in  P_{\alpha_2}, \mbox{ or vice versa}. 
\label{equation:refinement}
\end{equation}

The purpose of the above requirement is to define seed letters using a 
biologically significant hierarchical clustering of amino acids represented
by a tree. In Section~\ref{subsection:prebuilded-tree}, we will use a 
pre-defined hierarchical clustering to design efficient seed alphabets.
Then in Section~\ref{subsection:ab-initio-clustering}, we construct our
own clustering based on appropriate background and
  foreground models of amino acids distribution. Finally, in
  Section~\ref{subsection:non-hierarchical} we lift condition
  (\ref{equation:refinement}) and study ``non-hierarchical'' seed
  alphabets. 
\subsection{Transitive alphabets based on a pre-defined clustering}
\label{subsection:prebuilded-tree}
\begin{figure}[htb]\center
\vspace{-0.0cm}
\includegraphics[width=0.8\textwidth]{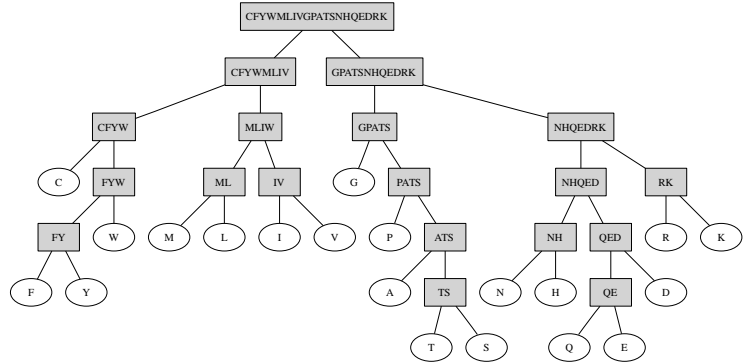}
\vspace{-0.1cm}
\caption{Hierarchical tree derived from~\cite{LiFanWangWangJPE03}.}
\label{figure:tree1}
\vspace{-0.0cm}
\end{figure}

Assume we have a biologically significant hierarchical clustering tree which is a
rooted binary tree $T$ with $20$ leaves labelled by amino acids. Such
trees have been proposed in
\cite{LiFanWangWangJPE03,MurphyWallqvistLevyJPE00}, based on different
similarity relations. 
The hierarchical tree derived from \cite{LiFanWangWangJPE03} is shown on 
Figure~\ref{figure:tree1}.
The tree, obtained with a purely bioinformatics analysis, groups
together amino acids with similar biochemical properties, such as
hydrophobic amino acids {\tt L,M,I,V}, hydrophobic aromatic amino
acids {\tt F,Y,W}, alcohols {\tt S,T}, or charged/polar amino acids
{\tt E,D,N,Q}. A similar grouping is obtained in
\cite{MurphyWallqvistLevyJPE00}. 

A {\em seed letter} is defined here as a subset $\alpha$ of nodes of $T$
such that
\begin{itemize}
\item[(i)] $\alpha$ contains all leaves,
\item[(ii)] for a node $v$, if $v\in\alpha$, then all descendants of
  $v$ belong to $\alpha$ too.
\end{itemize}
In other words, a seed letter can be thought of as a ``horizontal
cut'' of the tree. For example, for the tree of Figure~\ref{figure:tree1}
there are  $1597$ different seed letters. 
Seed letters are naturally ordered by inclusion.
The smallest one is the ``identity'' seed letter $\#$,
containing only the leaves.
The largest one is the ``joker'' seed letter $\_$,
containing all the nodes of $T$.
One particular seed letter is obtained by removing from
$\_$ the root node. We denote it by $@$.

Observe that each seed letter $\alpha$ represents naturally
an equivalence relation on $\Sigma$:
$a_i$ and $a_j$ are related iff their common ancestor belongs to
$\alpha$. It is identity relation in case of $\#$ and
full relation in case of $\_$. 

Following condition (\ref{equation:refinement}), a \emph{hierarchical seed alphabet} is a family $\G$ of seed letters
such that 
\begin{equation}
\mbox{for every }\al_1, \al_2 \in \G,
\mbox{ either }\al_1 \subseteq \al_2\mbox{ or }\al_2 \subseteq \al_1. 
\label{equation:consistent}
\end{equation}
Hence, a seed alphabet is a chain in the inclusion
ordering of seed letters. Let us analyze what are the maximal seed alphabets
wrt.\ inclusion. Clearly each maximal seed alphabet $\G$ always
contains the smallest and the largest seed letters $\#$ 
and $\_$.
Interestingly, each maximal $\G$ contains also $@$,
as $@$ is comparable (by inclusion) to any other seed letter. 

It can be shown that any maximal seed alphabet contains exactly $20$
letters that can be obtained by a stepwise merging of two subtrees
rooted at immediate descendants of some node $v$ into the subtree
rooted at $v$. Therefore, since a binary tree with $n$ leaves contains
$n-1$ internal nodes, a maximal seed alphabet contains precisely $20$
letters and can be specified by a permutation of internal nodes in
tree $T$. 
\subsubsection{Resulting alphabet}
Figure~\ref{figure:pre-defined-alphabet} shows alphabet 
{\polish} designed through the approach 
of this Section. The alphabet has been designed from the tree of
Figure~\ref{figure:tree1}. Each line corresponds to a letter (amino
acid partition). Among the alphabets obtained with different parameter
values, alphabet {\polish} produced better seeds and will be used in
the experimental part of this work (Section~\ref{section:experiments}). 
\begin{figure}[htb]
\vspace{-0.4cm}
{\tiny
\begin{eqnarray*}
& \{CFYWMLIVGPATSNHQEDRK\} \\[-0.6mm]
& \{CFYWMLIV\}\;        \{GPATSNHQEDRK\}\;\\[-0.6mm]
& \{CFYWMLIV\}\;        \{GPATS\}\;   \{NHQEDRK\}\;\\[-0.6mm]
& \{CFYW\}\;    \{MLIV\}\;    \{GPATS\}\;   \{NHQEDRK\}\;\\[-0.6mm]
& \{CFYW\}\;    \{MLIV\}\;    \{G\}\;       \{PATS\}\;    \{NHQEDRK\}\; \\[-0.6mm]
& \{C\}\;       \{FYW\}\;     \{MLIV\}\;    \{G\}\;       \{PATS\}\;    \{NHQEDRK\}\; \\[-0.6mm]
& \{C\}\;       \{FYW\}\;     \{MLIV\}\;    \{G\}\;       \{P\}\;       \{ATS\}\;     \{NHQEDRK\}\; \\[-0.6mm]
& \{C\}\;       \{FY\}\;      \{W\}\;       \{MLIV\}\;    \{G\}\;       \{P\}\;       \{ATS\}\;     \{NHQEDRK\}\;   \\[-0.6mm]     
& \{C\}\;       \{F\}\;       \{Y\}\;       \{W\}\;       \{MLIV\}\;    \{G\}\;       \{P\}\;       \{ATS\}\;     \{NHQEDRK\}\;     \\[-0.6mm]          
& \{C\}\;       \{F\}\;       \{Y\}\;       \{W\}\;       \{MLIV\}\;    \{G\}\;       \{P\}\;       \{A\}\;        \{TS\}\;      \{NHQEDRK\}\; \\[-0.6mm]
& \{C\}\;       \{F\}\;       \{Y\}\;       \{W\}\;       \{MLIV\}\;    \{G\}\;       \{P\}\;       \{A\}\;       \{T\}\;       \{S\}\;       \{NHQEDRK\}\;\\[-0.6mm]
& \{C\}\;       \{F\}\;       \{Y\}\;       \{W\}\;       \{MLIV\}\;    \{G\}\;       \{P\}\;       \{A\}\;       \{T\}\;       \{S\}\;       \{NHQED\}\;   \{RK\}\;  \\[-0.6mm]    
& \{C\}\;       \{F\}\;       \{Y\}\;       \{W\}\;       \{MLIV\}\;    \{G\}\;       \{P\}\;       \{A\}\;       \{T\}\;       \{S\}\;       \{NHQED\}\;   \{R\}\;       \{K\}\;    \\[-0.6mm]   
& \{C\}\;       \{F\}\;       \{Y\}\;       \{W\}\;       \{MLIV\}\;    \{G\}\;       \{P\}\;       \{A\}\;       \{T\}\;       \{S\}\;       \{NH\}\;      \{QED\}\;     \{R\}\;       \{K\}\;       \\[-0.6mm]
& \{C\}\;       \{F\}\;       \{Y\}\;       \{W\}\;       \{MLIV\}\;    \{G\}\;       \{P\}\;       \{A\}\;       \{T\}\;       \{S\}\;       \{N\}\;       \{H\}\;       \{QED\}\;     \{R\}\;       \{K\}\;    \\[-0.6mm]
& \{C\}\;       \{F\}\;       \{Y\}\;       \{W\}\;       \{MLIV\}\;    \{G\}\;       \{P\}\;       \{A\}\;       \{T\}\;       \{S\}\;       \{N\}\;       \{H\}\;       \{QE\}\;      \{D\}\;       \{R\}\;       \{K\}\;       \\[-0.6mm]
& \{C\}\;       \{F\}\;       \{Y\}\;       \{W\}\;       \{MLIV\}\;    \{G\}\;       \{P\}\;       \{A\}\;       \{T\}\;       \{S\}\;       \{N\}\;       \{H\}\;       \{Q\}\;       \{E\}\;       \{D\}\;       \{R\}\;       \{K\}\;       \\[-0.6mm]
& \{C\}\;       \{F\}\;       \{Y\}\;       \{W\}\;       \{ML\}\;      \{IV\}\;      \{G\}\;       \{P\}\;       \{A\}\;       \{T\}\;       \{S\}\;       \{N\}\;       \{H\}\;       \{Q\}\;       \{E\}\;       \{D\}\;       \{R\}\;       \{K\}\;       \\[-0.6mm]
& \{C\}\;       \{F\}\;       \{Y\}\;       \{W\}\;       \{M\}\;       \{L\}\;       \{IV\}\;      \{G\}\;       \{P\}\;       \{A\}\;       \{T\}\;       \{S\}\;       \{N\}\;       \{H\}\;       \{Q\}\;       \{E\}\;       \{D\}\;       \{R\}\;       \{K\}\;       \\[-0.6mm]
& \{C\}\;       \{F\}\;       \{Y\}\;       \{W\}\;       \{M\}\;       \{L\}\;       \{I\}\;       \{V\}\;       \{G\}\;       \{P\}\;       \{A\}\;       \{T\}\;       \{S\}\;       \{N\}\;       \{H\}\;       \{Q\}\;       \{E\}\;       \{D\}\;       \{R\}\;       \{K\}\;       \\[-0.6mm]
\end{eqnarray*}
}
\vspace{-1.0cm}
\caption[Alphabet based on
  Figure~\ref{figure:tree1}]{Alphabet {\polish} designed using the tree of
  Figure~\ref{figure:tree1}. Each line corresponds to a seed letter (amino acid partition)}
\label{figure:pre-defined-alphabet}
\vspace{-0.8cm}
\end{figure}
\subsection{Transitive alphabets using an {\em ab initio} clustering method}
\label{subsection:ab-initio-clustering}
\subsubsection{Hierarchical clustering of amino acids}
\label{subsubsection:ab-initio-clustering-algorithm}
A prerequisite to the approach of
Section~\ref{subsection:prebuilded-tree} is a given tree
describing a hierarchical clustering of amino acid based on some
similarity measure. In this section, we describe an {\em ab initio}
approach that constructs a hierarchical clustering of amino acids from
scratch, using a likelihood measure. 

As in Section~\ref{section:transitive-seed-alphabet}, our goal here is
to construct a family of seed letters verifying
(\ref{equation:consistent}). 
This family will be obtained with a simple greedy neighbor-joining
clustering algorithm, starting with the family of twenty amino acid singletons. 

We start with the partition of amino acids into 20 singletons. This
partition corresponds to the $\#$ letter. For a current partition 
$P = \{C_1,\ldots,C_n\}$, iteratively apply the following procedure.
\newpage
\begin{enumerate}
\item[1] For each pair of sets $C_k$, $C_\ell$,
  \begin{enumerate}
  \item[1.1] consider the set 
    $Bridge(C_k, C_\ell) = \{(a_i,a_j) | a_i \in C_k,\ a_j \in C_\ell\}$.
  \item[1.2] compute $ForeProb(k,\ell)=\sum\{f_{ij}|a_i\in C_k,\ a_j
    \in C_\ell\}$\\ and {~~~~~} $BackProb(k,\ell)=\sum\{b_{i}b_j|a_i\in C_k,\ a_j
    \in C_\ell\}$,
  \item[1.3] compute $L(k,\ell) =  ForeProb(k,\ell) / BackProb(k,\ell)$ 
  \end{enumerate}
\item[2] Find the pair of sets $(C_k, C_\ell)$ yielding the maximal
  $L(k,\ell)$,
\item[3] Merge $C_k$ and $C_\ell$ into a new set, obtaining a new
  partition. 
\end{enumerate}

The rationale behind this simple procedure is that those two sets of
amino acid are merged together which produce the maximal increment in
the likelihood $f(\alpha)/b(\alpha)$. An alternative method, when the likelihood of the whole
resulting set is maximized, yields biased results, as sets with a high
likelihood tend to ``absorb'' other sets. 
\subsubsection{Resulting alphabet}
\label{subsubsection:ab-initio-clustering-example}
An alphabet, called {\russianT}, obtained with this greedy
neighbor-joining approach is given in
Figure~\ref{figure:amino_acids_neigboor_joining}. It will be used in
experiments presented later in Section~\ref{section:experiments}. 
\begin{figure}\center
\vspace{-0.4cm}
{\tiny
\begin{eqnarray*}
& \{CFYWHMLIVPGQERKNDATS\}\; \\[-0.6mm]
& \{CFYWHMLIV\}\; \{PGQERKNDATS\}\; \\[-0.6mm]
& \{C\}\; \{FYWHMLIV\}\; \{PGQERKNDATS\}\; \\[-0.6mm]
& \{C\}\; \{FYWHMLIV\}\; \{P\}\; \{GQERKNDATS\}\; \\[-0.6mm]
& \{C\}\; \{FYWH\}\; \{MLIV\}\; \{P\}\; \{GQERKNDATS\}\; \\[-0.6mm]
& \{C\}\; \{FYWH\}\; \{MLIV\}\; \{P\}\; \{GATS\}\; \{QERKND\}\; \\[-0.6mm]
& \{C\}\; \{FYWH\}\; \{MLIV\}\; \{P\}\; \{G\}\; \{ATS\}\; \{QERKND\}\; \\[-0.6mm]
& \{C\}\; \{FYWH\}\; \{MLIV\}\; \{P\}\; \{G\}\; \{ATS\}\; \{QERK\}\; \{ND\}\; \\[-0.6mm]
& \{C\}\; \{FYW\}\; \{H\}\; \{MLIV\}\; \{P\}\; \{G\}\; \{ATS\}\; \{QERK\}\; \{ND\}\; \\[-0.6mm]
& \{C\}\; \{FYW\}\; \{H\}\; \{MLIV\}\; \{P\}\; \{G\}\; \{A\}\; \{TS\}\; \{QERK\}\; \{ND\}\; \\[-0.6mm]
& \{C\}\; \{FYW\}\; \{H\}\; \{MLIV\}\; \{P\}\; \{G\}\; \{A\}\; \{TS\}\; \{QE\}\; \{RK\}\; \{ND\}\; \\[-0.6mm]
& \{C\}\; \{FYW\}\; \{H\}\; \{ML\}\; \{IV\}\; \{P\}\; \{G\}\; \{A\}\; \{TS\}\; \{QE\}\; \{RK\}\; \{ND\}\; \\[-0.6mm]
& \{C\}\; \{FYW\}\; \{H\}\; \{ML\}\; \{IV\}\; \{P\}\; \{G\}\; \{A\}\; \{TS\}\; \{QE\}\; \{RK\}\; \{N\}\; \{D\}\; \\[-0.6mm]
& \{C\}\; \{FYW\}\; \{H\}\; \{ML\}\; \{IV\}\; \{P\}\; \{G\}\; \{A\}\; \{T\}\; \{S\}\; \{QE\}\; \{RK\}\; \{N\}\; \{D\}\; \\[-0.6mm]
& \{C\}\; \{FY\}\; \{W\}\; \{H\}\; \{ML\}\; \{IV\}\; \{P\}\; \{G\}\; \{A\}\; \{T\}\; \{S\}\; \{QE\}\; \{RK\}\; \{N\}\; \{D\}\; \\[-0.6mm]
& \{C\}\; \{FY\}\; \{W\}\; \{H\}\; \{ML\}\; \{IV\}\; \{P\}\; \{G\}\; \{A\}\; \{T\}\; \{S\}\; \{Q\}\; \{E\}\; \{RK\}\; \{N\}\; \{D\}\; \\[-0.6mm]
& \{C\}\; \{FY\}\; \{W\}\; \{H\}\; \{M\}\; \{L\}\; \{IV\}\; \{P\}\; \{G\}\; \{A\}\; \{T\}\; \{S\}\; \{Q\}\; \{E\}\; \{RK\}\; \{N\}\; \{D\}\; \\[-0.6mm]
& \{C\}\; \{FY\}\; \{W\}\; \{H\}\; \{M\}\; \{L\}\; \{I\}\; \{V\}\; \{P\}\; \{G\}\; \{A\}\; \{T\}\; \{S\}\; \{Q\}\; \{E\}\; \{RK\}\; \{N\}\; \{D\}\; \\[-0.6mm]
& \{C\}\; \{F\}\; \{Y\}\; \{W\}\; \{H\}\; \{M\}\; \{L\}\; \{I\}\; \{V\}\; \{P\}\; \{G\}\; \{A\}\; \{T\}\; \{S\}\; \{Q\}\; \{E\}\; \{RK\}\; \{N\}\; \{D\}\; \\[-0.6mm]
& \{C\}\; \{F\}\; \{Y\}\; \{W\}\; \{H\}\; \{M\}\; \{L\}\; \{I\}\; \{V\}\; \{P\}\; \{G\}\; \{A\}\; \{T\}\; \{S\}\; \{Q\}\; \{E\}\; \{R\}\; \{K\}\; \{N\}\; \{D\}\; \\[-0.6mm]
\end{eqnarray*}
}
\vspace{-1.0cm}
 \caption{Alphabet {\russianT} obtained with the method of Section
 \ref{subsubsection:ab-initio-clustering-algorithm}}  
 \label{figure:amino_acids_neigboor_joining}
\vspace{-0.8cm}
\end{figure}
\subsection{Non-hierarchical alphabets}
\label{subsection:non-hierarchical}
Previous approaches (Sections~4.1 and 4.2) were based on 
requirement (\ref{equation:consistent}) specifying that letters of the seed
alphabet should be embedded one into another to form a ``nested''
hierarchy. This requirement is biologically motivated and, on the
other hand, computationally useful as it reduces considerably the
space of possible letters. However, this requirement is not necessary
to implement the direct indexing (see Introduction). Therefore, we
also designed non-hierarchical alphabets in order to compare them to
hierarchical ones. We used the following heuristic consisting in 
generating first a large number of seed candidates, and selecting the
ones with (1) high likelihood ratio, (2) a range of different weights.
\subsubsection{Resulting alphabet}

An alphabet obtained with the above heuristic, called {\russianNT}, 
is shown in Figure~\ref{alph-non-hierar}. 
This alphabet will be used in the experiments reported in
Section~\ref{section:experiments}. 

\begin{figure}[htb]\center
\vspace{-0.4cm}
{\tiny
\begin{eqnarray*}
& \{ARNDCQEGHILMKFPSTWYV\}\;\\[-0.6mm]
& \{ARNDQEGHILMKFPSTWYV\}\; \{C\}\;\\[-0.6mm] 
& \{ARNDCQEHILMKFPSTWYV\}\; \{G\}\;\\[-0.6mm] 
& \{ARNDQEHILMKFSTYV\}\; \{CGPW\}\;\\[-0.6mm]
& \{ARCQEHILMKFSTYV\}\; \{NDGPW\}\;\\[-0.6mm]
& \{ARNDCQEGHKPST\}\; \{ILMFWYV\}\;\\[-0.6mm]
& \{ARNDQEGHKST\}\; \{CILMFWYV\}\; \{P\}\;\\[-0.6mm]
& \{ARNDQEHKPST\}\; \{CW\}\; \{G\}\; \{ILMFYV\}\;\\[-0.6mm]
& \{ARNDQEKST\}\; \{CP\}\; \{GHW\}\; \{ILMFYV\}\;\\[-0.6mm]
& \{AGPST\}\; \{RNDQEHK\}\; \{C\}\; \{ILMFWYV\}\;\\[-0.6mm]
& \{APST\}\; \{RNDQEHK\}\; \{CW\}\; \{G\}\; \{ILMFYV\}\;\\[-0.6mm]
& \{AGST\}\; \{RNDQEK\}\; \{C\}\; \{HFWY\}\; \{ILMV\}\; \{P\}\;\\[-0.6mm]
& \{AST\}\; \{RNDQEK\}\; \{CH\}\; \{G\}\; \{ILMV\}\; \{FWY\}\; \{P\}\;\\[-0.6mm]
& \{AST\}\; \{RQEHK\}\; \{ND\}\; \{CP\}\; \{G\}\; \{ILMV\}\; \{FWY\}\;\\[-0.6mm]
& \{AST\}\; \{RQK\}\; \{NH\}\; \{DE\}\; \{C\}\; \{G\}\; \{ILMV\}\; \{FWY\}\; \{P\}\;\\[-0.6mm]
& \{A\}\; \{RQK\}\; \{N\}\; \{DE\}\; \{C\}\; \{G\}\; \{H\}\; \{ILMV\}\; \{FY\}\; \{P\}\; \{ST\}\; \{W\}\;\\[-0.6mm]
& \{A\}\; \{RK\}\; \{N\}\; \{DE\}\; \{C\}\; \{QH\}\; \{G\}\; \{ILV\}\; \{M\}\; \{FY\}\; \{P\}\; \{ST\}\; \{W\}\;\\[-0.6mm]
& \{A\}\; \{RQK\}\; \{ND\}\; \{C\}\; \{E\}\; \{G\}\; \{H\}\; \{IV\}\; \{LM\}\; \{FWY\}\; \{P\}\; \{ST\}\;\\[-0.6mm]
& \{A\}\; \{RK\}\; \{ND\}\; \{C\}\; \{Q\}\; \{E\}\; \{G\}\; \{H\}\; \{IV\}\; \{LM\}\; \{FWY\}\; \{P\}\; \{S\}\; \{T\}\;\\[-0.6mm]
& \{A\}\; \{RK\}\; \{N\}\; \{D\}\; \{C\}\; \{Q\}\; \{E\}\; \{G\}\; \{H\}\; \{IV\}\; \{L\}\; \{M\}\; \{FY\}\; \{P\}\; \{S\}\; \{T\}\; \{W\}\;\\[-0.6mm]
& \{A\}\; \{R\}\; \{N\}\; \{D\}\; \{C\}\; \{QE\}\; \{G\}\; \{H\}\; \{I\}\; \{L\}\; \{K\}\; \{M\}\; \{FWY\}\; \{P\}\; \{S\}\; \{T\}\; \{V\}\;\\[-0.6mm]
& \{A\}\; \{R\}\; \{N\}\; \{D\}\; \{C\}\; \{Q\}\; \{E\}\; \{G\}\; \{H\}\; \{I\}\; \{L\}\; \{K\}\; \{M\}\; \{F\}\; \{P\}\; \{S\}\; \{T\}\; \{W\}\; \{V\}\;\\[-0.6mm]
\end{eqnarray*}
}
\vspace{-1.0cm}
\caption[Non-hierarchical alphabet]{Non-hierarchical alphabet {\russianNT}.}
\label{alph-non-hierar}
\vspace{-0.4cm}
\end{figure}
\section{Experiments}
\label{section:experiments}
This section describes the experiments we made to
test the efficiency of seeds we designed with different
methods of previous sections. 
Sections~\ref{subsection:methods}-\ref{blast-evaluation} describe
the experimental protocol, from the assignment of background
and foreground probabilities, to the seed design. In
Section~\ref{subsection:results_theoretical}, we analyze the power of
different seed models proposed in
Sections~\ref{section:non-transitive-seed-alphabet}-\ref{section:transitive-seed-alphabet}
with respect to probabilistic models. 
\subsection{Probability assignment and alphabet generation}
\label{subsection:methods}
First of all, we derived probabilistic models in accordance with the
BLOSUM62 data from the original paper \cite{BLOSSUM92}. We obtained the
BLOCKS database (version 5) \cite{BLOCKS91} and the software of
\cite{BLOSSUM92} to infer Bernoulli probabilities for the background
and foreground alignment models. These probabilities have been used
throughout the whole pipeline of experiments. 

Different seed alphabets have then been generated by the methods 
presented in Section~\ref{section:non-transitive-seed-alphabet}
(alphabet \nontrans),
Section~\ref{subsection:prebuilded-tree} (alphabet \linebreak[4]\polish),
Section~\ref{subsection:ab-initio-clustering} (alphabet \russianT) and
Section~\ref{subsection:non-hierarchical} (alphabet \russianNT). 
\subsection{Seed design}
\label{seed-design}
To each alphabet, we applied a seed design procedure that we briefly
describe now. 
Since each seed (or seed family) is characterized by two parameters 
-- sensitivity and selectivity -- 
it can be associated with a point on a 2-dimensional 
plot. Best seeds are then defined to be those which belong to
the {\em Pareto} set among all seeds, i.e. those than cannot be
strictly improved by increasing sensitivity, selectivity, or both. 

For different selectivity levels, we designed good seed families
containing one to six individual seeds, among which the best family was
selected. In each seed family, each individual seed has been
assumed to have approximately the same weight, within 5\%
tolerance. This requirement is natural as in the case of
divergent weights, seeds with lower weight would dominantly affect the
performance. In practice, having individual seeds of similar
weight allows an efficient parallel implementation
(see e.g. \cite{PeterlongoEtAlPBC07}). 

Estimation of sensitivity of individual seeds or seed families has
been done with the algorithm described in
\cite{KucherovNoeRoytberg06} and implemented in the {\sc Iedera}
software, available at {\tt {\small
    \url{http://bioinfo.lifl.fr/yass/iedera.php}}}. 
The selectivity of an individual seed has been computed according to
the definition (Section~\ref{section:preliminaries}). For a seed
family, its selectivity has been lower-estimated by summing the
background probabilities of individual seeds. 

Seed family design has been done using a hill climbing heuristics (see
\cite{BuhlerKeichSunRECOMB03,IlieIlieBIOCOMP07,ZhouStantonFloreaBMCBioinformatics08}) alternating seed generation and seed estimation steps. 
All experiments were conducted for alignment lengths 16 and 32. 
\subsection{{\sc Blastp} and the vector  seed family from \cite{BrownTCBB05}}
\label{blast-evaluation}
Our goal is to compare between different seed design approaches
proposed in this paper, but also to benchmark them against other
reference seeding methods. We used two references: the {\sc Blastp} 
seeding method and the family of vector seeds proposed in
\cite{BrownTCBB05}. Both of them use a score (or weight) resulting
from the accumulative contribution of several neighboring positions to
define a hit (see Introduction). Therefore, they use a more powerful
(and also more costly to implement) formalism of seeding. 

To estimate the sensitivity and selectivity of those seeds, we
modified our methods described in the previous section by representing
an alignment by a sequence of possible individual scores. Foreground
and background probability of each score is easily computed from those
for amino acid pairs. After that,
sensitivity and selectivity is computed similarly to the previous
case. 
\subsection{Results}
\label{subsection:results_theoretical}
\begin{figure}[p]\center
\vspace{-0.1cm}
\caption[ROC curves on theoretical
  models]{\label{figure:ROC-theoretical-zoom}ROC curves of seed
  performance measured on the probabilistic model}
\vspace{-0.1cm}
\includegraphics[width=12.5cm]{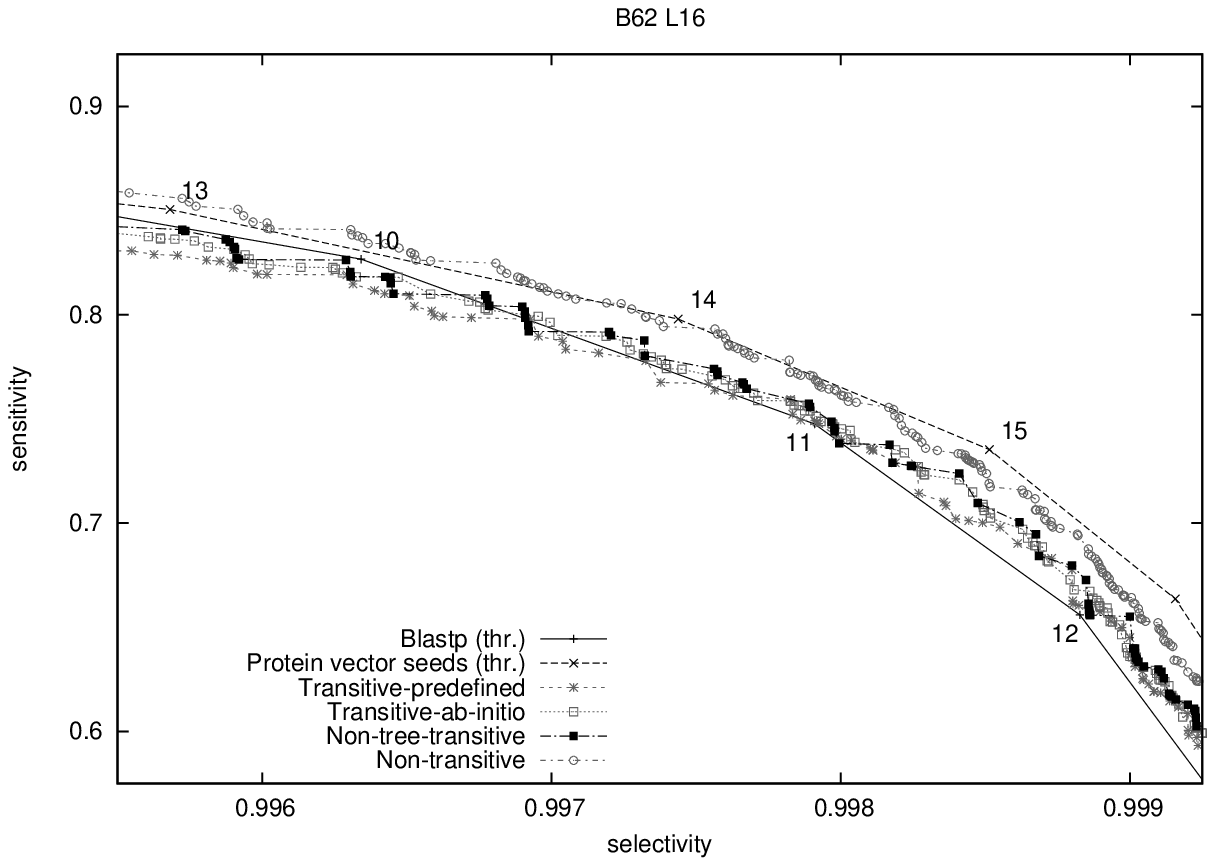}
\includegraphics[width=12.5cm]{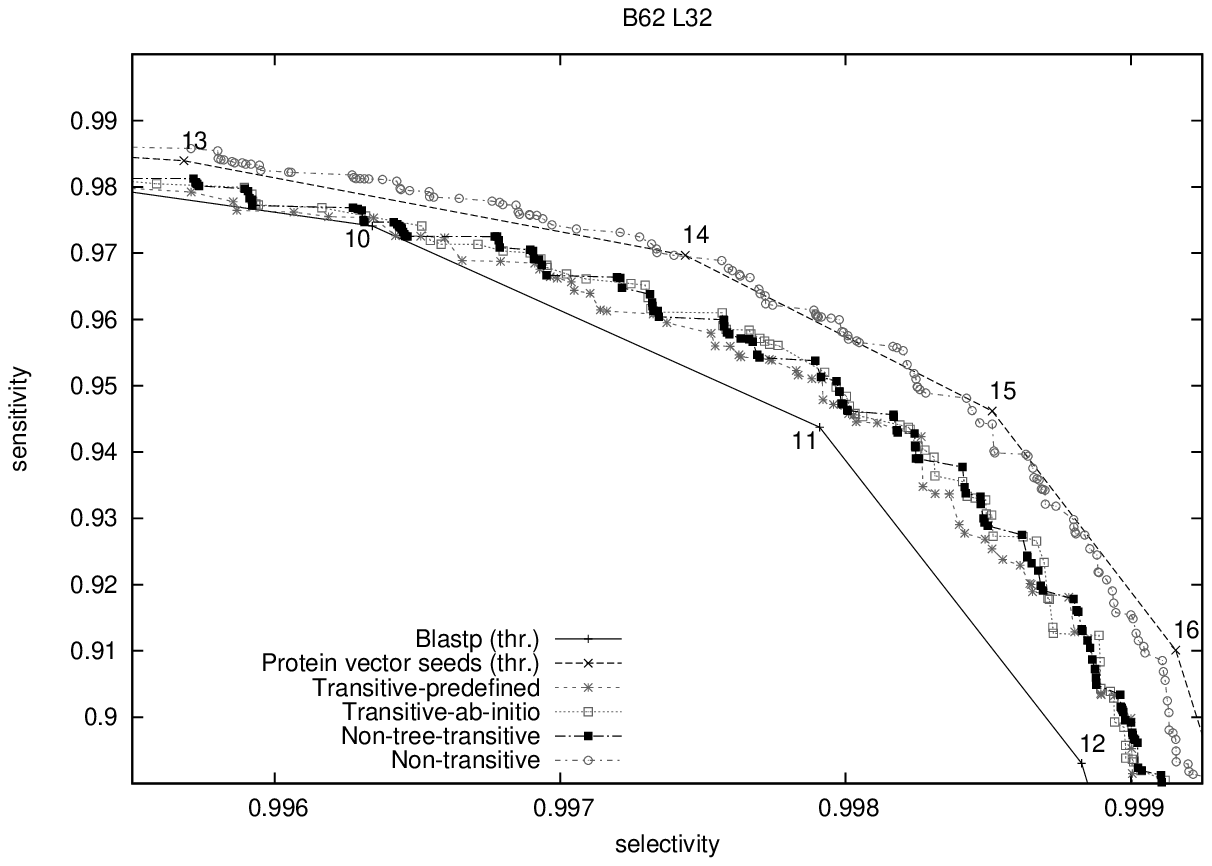}
\vspace{-0.1cm}
\label{ROC-curves}
\end{figure}

We compare the performance of the different approaches by
plotting ROC curves of Pareto-optimal sets of seeds on the
selectivity/sensitivity graph. The two plots in Figure
\ref{figure:ROC-theoretical-zoom} show the results for alignment
length 16 and 32 respectively. The two first polylines show the 
performance of {\sc Blastp} with word size 3 and the vector seed
family from \cite{BrownTCBB05}, for different score thresholds. The
other curves show the performances of different seed alphabets
from
Sections~\ref{section:non-transitive-seed-alphabet}-\ref{section:transitive-seed-alphabet}
represented by the Pareto-optimal seeds (seed families) that we were able to
construct over those alphabets. As mentioned earlier in
Section~\ref{seed-design}, each time we selected the best seed family
among those with different number of individual seeds. 
Typically (but not exclusively), points on the plots correspond to
seed families with 3 to 5 seeds.
Typically, the seed span ranges between 3 and 5 (respectively, 3 and 6) 
for alignment length 16 (respectively, 32). 
Seeds with langer span ($>4$) tend to occur in seed families with larger
number of seeds ($>3$). 

We observe that non-transitive seeds over the alphabet of
Section~\ref{section:non-transitive-seed-alphabet}
are comparable in performance with the vector seed
family from \cite{BrownTCBB05} and clearly outperforms seeds over
other alphabets. This result is interesting in itself, although this
alphabet is unpractical in many cases, due to its incompatibility with
the transitivity condition. 

As for the other alphabets, they roughly show a comparable
performance among them. Note that using non-hierarchical alphabet
(Section~\ref{subsection:non-hierarchical}) does not bring much of
improvement, which justifies condition (\ref{equation:consistent}). 
For the alignment length 16, our seeds perform comparably to
{\sc Blastp}, with a slightly better performance for high thresholds
and a slightly worse performance for low thresholds. On the other
hand, for alignments of length 32, our seeds clearly outperform {\sc
  Blastp}.
\section{Conclusion}

The main conclusion of our work is that although the subset seed model
is less expressive than the method of accumulative score used in 
{\sc Blastp}, carefully designed subset seeds can reach the same or
even a higher performance. To put it informally, the use of the
accumulative score in defining a hit can, without loss of performance,
be replaced by a careful distinction between different amino acid
matches without using any scoring system. From a practical point of 
view, subset seeds can provide a more efficient implementation,
especially for large-scale protein comparisons, due to
a much smaller number of accesses to the hash table. In particular,
this can be very useful for parallel implementations or specialized
hardware (see e.g. \cite{PeterlongoEtAlPBC07}). 

Note that the seed design heuristic sketched in
Section~\ref{seed-design} does not guarantee to compute optimal
seeds, and therefore our seeds could potentially be further improved
by a more advanced design procedure, possibly bringing a further
increase in performance. This is especially true for seeds of large
weight (due to a bigger number of those), for which our seed design
procedure could produce non-optimal seeds, thus explaining some
``drop-offs'' in high-selectivity parts of plots of
Figure~\ref{ROC-curves}.  

As far as further research is concerned, the question of efficient
seed design remains an open issue. Improvements of the hill climbing
heuristics used in this work are likely to be possible. 

\paragraph{Acknowledgements}
Parts of this work have been done during visits to LIFL of Ewa
Szczurek (June-August 2006), Anna Gambin and S\l{}awomir Lasota
(August 2006) and Mikhail Roytberg (October-December 2006). Those
visits were supported by the ECO-NET and Polonium programs of the
French Ministry of Foreign Affairs.

\bibliographystyle{splncs}
\bibliography{paper}
\end{document}